\documentclass[11pt,titlepage]{article}

\usepackage{setspace} 

\usepackage{amsmath}

\usepackage{graphicx}

\usepackage[round,numbers,sort&compress]{natbib} 
\usepackage{amssymb,amsmath,amsfonts}
\usepackage{graphicx}
\usepackage{units}
\usepackage{caption}

\graphicspath{{Figures/}}

\title{Branching actin network remodeling governs the force-velocity
  relationship.}

\author{Daniel B. Smith \\
  Department of Mathematics \\
  University of Pittsburgh, Pittsburgh, PA \\
  National Heart, Lung, and Blood Institute \\
  National Institutes of Health, Bethesda, MD
  \and Jian Liu\thanks{
    Corresponding author.  Address: 
    National, Heart, Lung and Blood Institute,
    National Institutes of Health,
    Building 50 - Louis B Stokes Lab, 3306,
    Bethesda, MD~20892, U.S.A.,
    Tel.:~(301)594-5862, Fax:~(301)402-3405} \\
  National Heart, Lung, and Blood Insitute, \\
  National Intistutes of Health, Bethesda, MD}

\date{}

\begin{document}

\maketitle

\abstract{Actin networks, acting as an engine pushing against an external load,
  are fundamentally important to cell motility. A measure of the effectiveness
  of an engine is the velocity the engine is able to produce at a given force,
  the force-velocity curve. One type of force-velocity curve, consisting of a
  concave region where velocity is insensitive to increasing force followed by
  a decrease in velocity, is indicative of an adaptive response. In contrast,
  an engine whose velocity rapidly decays as a convex curve in response to
  increasing force would indicate a lack of adaptive response. Even taken
  outside of a cellular context, branching actin networks have been observed to
  exhibit both concave and convex force-velocity curves. The exact mechanism
  that can explain both force-velocity curves is not yet known. We carried out
  an agent-based stochastic simulation to explore such a mechanism. Our results
  suggest that upon loading, branching actin networks are capable of remodeling
  by increasing the number filaments growing against the load. Our model
  provides a mechanism that can account for both convex and concave
  force-velocity relationships observed in branching actin networks. Finally,
  our model gives a potential explanation to the experimentally observed force
  history dependence for actin network velocity.

}

\clearpage

\section*{Introduction}
Branching actin networks are the principle engine that drives cell motility
ranging from cell migration \cite{Pollard2003a, Fletcher2010} to endomembrane
trafficking \cite{Engqvist-Goldstein2003}. In the lamellipodium of migrating
cells, actin filaments grow from their barbed-ends, pushing against the plasma
membrane in the direction of cell movement. New filaments branch off of
existing filaments through the actin related protein (Arp2/3) complex,
activated by WASP at the membrane. Filaments branch at a characteristic angle
of $\sim70^\circ$. Capping proteins limit the growth of filaments by binding to
the barbed-end of the filament. At the back of the network, actin filaments
depolymerize and are severed, providing a fresh actin monomer supply to the
front \cite{Pollard2003a, Fletcher2010}. Understanding the basic process by
which an actin network is able to exert force against a load is a fundamental
step to understanding a number of cellular processes \cite{Pollard2009}.

Both in vitro and in vivo experiments have been performed to probe the
force-velocity relationship of growing actin networks \cite{McGrath2003,
  Prass2006, Heinemann2011, Cameron1999, Parekh2005, Marcy2004}. Migrating
cells show an adaptive response exhibiting a concave force-velocity
relationship \cite{Prass2006, Heinemann2011}. However, the concave
force-velocity relationship is preceded by a large reduction in velocity to
small forces \cite{Prass2006}. The mechanism controlling the concave
force-velocity relationship and the initial response to small forces in cells
is complicated by other cellular components such as focal adhesions. To study
the exact mechanism that determines the force-velocity relationship of a
branching actin network, in vitro experiments with more controllable conditions
have been performed. One study measured the velocity of an actin network
growing against a constant load force set by an atomic force microscope, and
the resulting force-velocity relationship was convex \cite{Marcy2004}. In a
different in vitro experiment, the actin network grew against the flexible
cantilever. The load force thus progressively increased as actin polymerization
deflected the cantilever, and the network showed a concave force-velocity
relationship \cite{Parekh2005}. That experiment also showed a hysteresis effect
where the velocity of the network was dependent upon the past forces applied to
the network.

Experiments done in vitro have demonstrated both convex and concave
force-velocity relationships in branching actin networks. This suggests that
actin networks can respond to external forces in both an adaptive and a
non-adaptive manner outside of cellular context. Even within the
simplified in vitro setting, it is still unclear how the individual factors
that govern branching actin network dynamics generate both the convex and the
concave curves. It has previously been proposed that the actin network remodels
itself in response to force \cite{Parekh2005}, but the nature of such
remodeling is largely unknown.

Evidence suggests that actin filaments utilize thermodynamic free energy to add
additional monomers to exert force towards the leading edge of the network
\cite{Shaevitz2007}. The proposed model for that behavior has been termed the
Brownian ratchet \cite{Peskin1993, Mogilner1996, Mogilner2003b}. The Brownian
ratchet mechanism takes advantage of the asymmetry in the on and off rates of
actin monomer binding to an existing filament. Small gaps arise between the
actin filaments and the leading edge due to thermodynamic
fluctuations. Monomers are able to bind during such fluctuations and push the
leading edge forward. The predicted force-velocity curve for a single filament
is a convex negative exponential function. When many filaments grow against the
same load, they are able to share the load.  A simple force-sharing
mechanism predicted that the force-velocity curve is nonetheless convex
\cite{Schaus2008}, even though its slope is shallower than that for single
filament. A model that tries to explain the force-insensitive region of the
concave force-velocity curve \cite{Carlsson2003} is largely incapable of
accounting for the convex curve.

Our theoretical model was built to study both types of force-velocity
relationships for branching actin networks. We used an agent-based stochastic
simulation method inspired by Weichsel et al. \cite{Weichsel2010} and Schaus
and Borisy \cite{Schaus2007}. Our results show that the balance between growth,
branching and capping events controls the ability of a branching actin network
to reinforce itself against a load. The model can explain both convex and
concave force-velocity relationships.

\subsection*{Model Description}
To discern the physical mechanism governing the force-velocity relationships of
branching actin networks, we aimed to construct the simplest model able to
reproduce the observed effects without compromising the physical essence of
actin networks. The model therefore only includes the four essential processes
of branching actin networks in \emph{in vitro} conditions. Below, we describe
the qualitative features of our model.

1.  Filaments grow by adding new monomers to the barbed end of the
filaments. When the filament is not in contact with the load, it does not feel
the load and, hence, it could grow at its free rate $V_0$. When a filament is
in contact with the load surface, the rate of growth follows the Brownian
ratchet mechanism, where the growth velocity of the filament is reduced by a
Boltzmann factor:
\begin{equation}\label{Brown}
  \mathcal{B}(F) = \exp\left[-\frac{F\delta}{k_BT}\right]
\end{equation}
where $F$ is the force felt by the filament, $\delta$ is the length of an
individual actin monomer, $k_B$ is Boltzmann’s constant, and $T$ is the
absolute temperature. 

2.  New filaments are created by branching off existing filaments. Arp2/3 binds
to existing filaments and creates a site for a new filament to grow and
generates the characteristic $70^\circ$ angle in between the original and newly
branched filaments \cite{Mullins1998}. Branching was modeled as a zero-order
reaction, independent of the number of actin filaments, which is consistent
with the experiments that suggest WASP/Scar-mediated Arp2/3 activation is the
limiting factor of network growth \cite{Pollard2000a}. And the branching angle
followed a Gaussian distribution that centers at $70^\circ$, with a standard
deviation of $5^\circ$ \cite{Mullins1998}.

3.  Capping proteins can bind to the tips of actin filaments, preventing them
from further elongation. In \emph{in vitro} conditions, the lifetime of barbed
end-bound capping proteins is $\sim$30 min \cite{Schafer1996}. This feature is
modeled by filament growth stopped once capped. We modeled filament capping as
a first-order reaction: the capping rate was proportional to the number of free
barbed ends, which is agreement with \emph{in vitro} experiment measurements
\cite{Mullins1999a, Xu1999}.

4.  A significant factor in the efficiency of a growing actin network against a
load is the ability of the network to share the load across multiple filaments
at the leading edge, which has been a recent topic of study
\cite{Schaus2008}. We implemented a similar load-sharing scheme to Schaus and
Borisy \cite{Schaus2008} among the filaments in contact with the load surface,
i.e., the sum of the load force felt by each contacting filament is in balance
with the total load. Note that the load forces felt by individual contacting
filaments across the leading edge are not equal, because the closer the
filament orientation is to the normal of the load surface, the larger share of
the load force this filament will feel.

At each time step, calculations were done in the following order. First, the
location of the leading edge was determined by the location of the foremost
filament. Then, filaments in contact with the leading edge (see methods for
definitions) were located. Filaments were next capped and branched, the
rates of which were calculated based on the location of the filaments and
Poisson statistics. Finally, the positions of all the filament tips were
updated by the growth rate for each filament in accordance to force-sharing
mechanism.

Model parameters were estimated based on experiment evidence where possible,
and they are listed in Table 1 with references. 

\section*{Methods}

\subsection*{Actin Model}
Each filament was modeled as a point in a two-dimensional plane representing
the filament barbed end. The plane was bounded by a hard leading edge in the
principle direction of motion (X) and periodic boundaries in the perpendicular
(Y) direction.  Each filament had three properties: an X-coordinate, a
Y-coordinate and an angle of growth $\theta$ which was relative to the
X-direction. As the filament grew, the X and Y coordinates would change in
time, e.g. $\dot{x}_j=v_j\cos(\theta_j)$ for filament $j$ growing at an angle
$\theta_j$, while the angle ($\theta$) for each filament did not change. The
filaments were limited to having angles in $-90^\circ < \theta < 90^\circ$ as
filaments growing against the principle direction of growth (X) are not seen in
experiment \cite{Mogilner2009a} and would quickly grow too far away from the
leading edge to contribute to the network velocity. The default parameters for
the model are listed in Table \ref{parameters}.

The filaments grew against the force exerted by the leading edge (at point
X=0). A filament was treated as being in contact with the boundary if the tip
of the filament was within one subunit length ($\delta$) from the leading edge.
Identifying the filaments that were in contact with the leading edge was
performed at the beginning of each time step.

Within the simulations, new filaments were generated by branching from existing
filaments. The filaments branched at a constant rate, calculated using
zeroth-order Poisson statistics. Having a constant rate of branching new
filaments essentially assumes that the rate limiting factor is the
concentration of Arp 2/3 \cite{Pollard2000a, Carlsson2003, Schaub2007a,
  Weichsel2010}. The initial tip of a new branch was placed at a point along
the initial filament randomly selected from a uniform distribution extending a
distance of $5\,\delta$ backwards from the tip of the initial filament. The
difference between the angle of the branching filament ($\theta_b$) and the
angle of the initial filament ($\theta_i$) was drawn from a Guassian
distribution ($\mathcal{N}$) with mean $70^\circ$ and variance
$\sigma^2=25^\circ$, i.e. $(\theta_b-\theta_i)\sim\mathcal{N}(70,25)$, and the
filaments branched in both directions ($\theta_b > \theta_i$ or $\theta_b <
\theta_i$) with equal probability \cite{Mullins1998}. Gaussian random numbers
were drawn using the gasdev algorithm in \cite{Press1992}. The filaments were
capped as a first-order reaction, with first-order Poisson statistics, meaning
the capping rate was dependent upon the number of active filaments not in
contact with the leading edge. Once a filament was capped, it was no longer
able to grow, branch or exert force upon the leading edge
\cite{Schafer1996}. The first-order statistics were calculated based on the
total number of active, uncapped filaments not in contact with the leading
edge. At each time step, the number of filaments added and capped during the
time were calculated using a uniform (0,1) random number using the ran1
algorithm in Numerical Recipes \cite{Press1992}. Filaments in contact with the
leading edge were neither branched nor capped as in previous models
\cite{Carlsson2001, Schaus2007, Schaus2008}.

Filaments grew at a deterministic, constant rate in the direction defined by
$\theta$, based on the assumption that the rate of actin subunits adding to the
filament is much faster than the rate of the network reorganization. We were
principally interested in the geometric structure of the network and considered
only average growth dynamics. Similar to \emph{in vitro} conditions, the actin
monomer concentration was modeled as saturated leading to a constant growth
rate of $100\,\delta/s$.  The velocity of each individual filament in contact
with the leading edge was reduced by the Boltzmann factor derived from the
Brownian-ratchet mechanism \cite{Peskin1993, Mogilner1996, Mogilner2003b}. The
force was shared across all filaments using a modified version of the optimal
force sharing from Schaus and Borisy \cite{Schaus2008}. Each time a filament
would add an additional actin subunit, the energy penalty would be proportional
to $F_{tot}\delta\cos(\theta)$, where $F$ is the total force applied by the
leading edge. Normalizing the individual forces so that the total force adds up
to $F_{tot}$ gives the final velocity expression for a filament $i$ in contact
with the leading edge:
\begin{equation}\label{velo}
  v_i = v_0 \exp\left(\frac{-F_{tot}\delta\cos(\theta_i)}
  {k_BT\sum_j\cos(\theta_j)}\right)
\end{equation}
where the sum $j$ is over all of the filaments in contact with the leading edge
and $F_{tot}$ is the total force applied to the system.

At each time step, calculations were done in the following order. First, the
location of the leading edge was calculated, which was defined as the largest X
coordinate of the active filaments. The filaments which were then within
$\delta$ of the leading edge are marked as in contact with the leading edge and
were no longer able to be capped or branch new filaments. Next, the number of
branching and capping events between times $t$ and $t+dt$ was calculated by
comparing a uniform random number in (0,1) and the cumulative distribution
function for the associated Poisson distribution. Individual filaments were
chosen randomly from the population of filaments not in contact with the
leading edge to be capped. Following that, filaments were again randomly chosen
to serve as the branching point for new filaments.  Next, the normalizing
constant was calculated by summing over the filaments in contact with the
boundary to get the average force exerted by each filament. Finally, the
positions of the filaments are advanced by $v_idt$ where $v_i$ is the velocity
of filament $i$.
  
All simulations started with independent, identically distributed initial
conditions. The simulations started with 200 filaments uniformly distributed in
the y direction and uniformly distributed within a box of length $20\,\delta$
in the X direction with the constraint that half of the filaments were in the
first $10\,\delta$ and half in the second $10\,\delta$. The initial filament
orientations were randomly drawn from a uniform distribution on
$(-90^\circ,90^\circ)$. The time step, $dt$, used in each simulation was
$10^-2$s.

\subsection*{Short Time-Scale}
One thousand simulations were run with independent, identically distributed
initial conditions as above. The simulations were averaged at each time step to
generate Figure 1(A). Figure 1(B) was generated by taking the minimum velocity
after force was applied. Parameter values were chosen to be $\kappa=1$ and
$\lambda=200$ to emphasize that we were only able to see convex force velocity
curves on short time-scales.

\subsection*{Equilibrium Simulations}
Each data point from Figure 2 was the result of averaging 10 equilibrium
simulations with constant force starting from the initial conditions described
above. Simulations were run for a total of 10,000 s with a $dt$ of $10^{-2}$ s
and data was sampled every 0.2 s. Reported data was sampled from the second
half of the simulation to allow the system to minimize the influence of initial
conditions. The equilibration can be seen in supplemental Figures S1-2.

\subsection*{Hysteresis Simulations}
The hysteresis simulations were performed with the same initial
conditions. Simulations were run for a total of 7,500 s.  The first 2,500 s
were run at an initial (low) force $f_0$ to equilibrate the system to the
velocity observed in Figure 2. At time 2,500 s, the force was increased to
$f_1$. The system was allowed to equilibrate again to a lower
velocity. Finally, at time 5,000 s the force was reduced back to $f_0$ where
the velocity rapidly rose before converging back to equilibrium.

Filaments were made to 'stick' to the leading edge by reducing the amount of
force felt by a filament close to $\delta$ away from the leading edge. The new
expression force felt by each filament was $F_j = F_j^\star q(x)$ where
$F_j^\star$ is the force used in \eqref{velo} and $q(x)$ is a cubic polynomial
such that:
\begin{equation}
  q(x) = 
  \begin{cases}
    1 & \text{if } 0\leq x\leq \gamma \\
    -2\left(\frac{1-x}{1-\gamma}\right)^3+3\left(\frac{1-x}{1-\gamma}\right)^2
    & \text{if } \gamma\leq x \leq 1 \\
    0 & \text{if } x>1
  \end{cases}
\end{equation}
where $\gamma=0.9$ for our simulations.

\subsection*{Curve Parameter}
We characterized the force-velocity curves by their $f_{1/2}$, the force at
which the relative velocity was reduced to $\frac{1}{2}$. Twenty constant force
simulations were performed for each pair of branching and capping rates ranging
from 850 pN to 17 nN. $f_{1/2}$ was estimated by linearly interpolating between
the two successive forces where the velocities surrounded $\frac{1}{2}$. When
the velocity at 17 nN was greater than one half, we used the largest force
calculated ($\sim$17 nN) for the purposes of figure \ref{phase}. Likewise, for
the purely convex curves with $f_{1/2}<4$ nN the value 4 nN was used in the
graph for clarity.

\section*{Results}

Our simulations were focused on how the collective properties of a branching
actin network influence the ability of the network to grow against a flat
surface applying a load force.

\subsection*{Fast force-velocity response is always convex}

The first set of simulations we ran tested the temporal response of the
velocity of branching actin network against a fixed load force. Figure 1(A)
shows that upon loading force, the velocity drops almost instantly, which then
recovers in a longer timescale ($\sim$minutes) reaching a value lower than
before the application of a load. Taking the velocity at the bottom of the
initial response to force response gives a force-velocity curve for the short
time-response of the network. Our model suggests that the force-velocity curve
is always convex (Figure 1(B)).

\subsection*{Force-velocity relationship at long-time response can be either 
  convex or concave}

Running the simulations for an extended amount of time allows us to study the
equilibrium force velocity relationship. Figure 2(A) shows that we were able to
reproduce both convex and concave force-velocity curves. The only difference
between the sets of simulations is the absolute value of the capping and
branching coefficients. Their ratio, and therefore the average number of
filaments, was fixed. We hypothesize that the network is able to reinforce
itself by bringing more filaments to the leading edge.

\subsection*{Branching actin networks remodel by increasing the number of 
  filaments contacting the leading edge}

An actin network can be visualized as a population of filaments contacting the
boundary and another population of filaments trailing the leading edge in
reserve. The network reinforces the filaments at the leading edge when trailing
filaments grow to reach the leading edge. That remodeling response simply
depends on the rate at which trailing filaments are able to catch up to the
leading edge. Figure 3 is an explanatory diagram showing a hypothetical
branching pattern. The first filament is in contact with the boundary, the
second filament branches off the first one, is further back, and serves as a
source for new filaments. We term such filaments “reserve filaments” (Figure
3(A)). Since these reserved filaments are not in contact with the load surface,
they don’t feel the load and grow at their free rate faster than the leading
edge. Consequently, some of the reserved filaments catch up to the leading
edge. When the capping rate is high, these reserved filaments will get capped
and stop growing before reaching the load surface (Figure 3(B)). Conversely,
when capping rate is low, these reserve filaments grow to be in contact with
the load surface, thereby reinforcing the leading edge against the load (Figure
3(C)).

The hypothesis that the network reinforces itself by filaments growing to the
leading edge would suggest that a network composed of shorter filaments, being
less likely to grow to the leading edge, would stall at lower
forces. Increasing the capping rate causes the filaments to grow for shorter
periods of time leading to shorter filaments, and indeed, Figure 2(B) shows
that by changing the capping rate, the force-velocity curve continuously
changes from concave to convex shapes. Fewer filaments reinforce the leading
edge leading to lower and lower stall forces. The number of filaments in
contact with the leading edge attests the level of reinforcement. Figure 2(C)
shows that more filaments are recruited to the leading edge when the capping
rate is low for the same level of force.

Figure 2(C) shows that the number of filaments in contact with the leading edge
increases with increasing force for all cases. However, the cases with lower
capping rates generate a larger increase before peaking and stalling. This is
because the branching rate is zeroth order, where the rate is independent of
the number of filaments; whereas the capping rate is first order, whose rate
increases with the number of available filaments. Before reaching the peak, the
network is, in some sense, branching dominated. On the other hand, increases in
force increase the number of filaments in contact with the leading edge, so
does the capping rate. When the effective capping rate becomes faster than that
of branching, the contacting filaments will get capped and, hence, their number
will drop off. It is also of note that the specific filaments in contact with
the leading edge are constantly changing. When the capping effect dominates
over branching, the leading edge velocity is no longer sufficient to provide
the turnover to sustain the higher number of filaments. Approximately at the
peak of Figure 2(C), the network begins to stall in the force-velocity curves
in Figure 2B.

\subsection*{Branching actin network remodeling could account for the observed 
  hysteresis effect}

The accumulation of filaments at the boundary is also able to explain some of
the hysteresis observed in actin networks. As seen in Figure 2, the number of
filaments in contact with the boundary increases with increasing
force. Subjecting the network to a large force and subsequently releasing that
force should leave excess filaments in contact with the leading edge.

Figure 4(A) shows the velocity of a simulation where the network pushed against
low force for the first third of the run, followed by high force in the middle
third, and finally the original low force. That is comparable to the experiment
in Parekh et al. \cite{Parekh2005}. The velocity initially shoots back up in
response to the reduced force, but it rapidly decays back to the initial
equilibrium force similar to \cite{Carlsson2003}. The reduction in the load
force speeds up the growth rate of all the contacting filaments at the leading
edge. Due to the angle dependence of the load sharing for individual filaments,
the speed-up of growth rates is heterogeneous across the contacting
filaments. As a result, some filaments grow faster while staying in con- tact
with the load, and slower growing filaments slide of the leading edge and are
capped. Ultimately, the number of contacting filaments relaxes back to the
velocity corresponding to the original force and completely loses its memory of
the previous loading force. Figure 4(B) shows that sustained hysteresis can be
realized in the model, if we incorporate a factor that causes the actin
filaments to stick to the leading edge. While we do not know the exact nature
of such an interaction between the filament tips and the load surface, actin
tethering has been theoretically proposed \cite{Mogilner2003b} and has some
experimental evidence \cite{Giardini2003, Upadhyaya2003a, Carlier2007a, Co2007,
  Svitkina2007}.

\subsection*{The balance between capping and branching events dictates actin 
  network remodeling}

Our results suggest that the actin network remodels itself by changing the
number of filaments in contact with the leading edge. That remodeling in turn
determines the shape of the force velocity relationship. In particular, it
determines the length of the concave portion of the curve. The balance between
the branching and capping rates controls the nature of the remodeling. To
obtain a systematic, systematic understanding of the rate dependence of the
actin force-velocity relationship, we carried out a phase diagram study for the
capping and branching rates. We used the force at which the velocity drops to
50\% of the small load force to characterize the shape of the force velocity
curve. Figure 5 shows the estimated $f_{1/2}$ values at each parameter value
that constitutes our principle prediction: faster capping rates lead to less
concave force velocity curves, and faster branching rates lead to more concave
curves.

\section*{Discussion}

We have proposed a simple mechanism where branching actin networks remodel
against a load force. The model shows that the initial response of branching
actin networks to loading always gives a convex force-velocity relationship
(Figure 1). On longer time scales, smaller capping rates and larger branching
rates generate more concave force-velocity relationship (Figure 2 and 5).

A number of recent theoretical studies have focused on how to explain both the
convex and concave force-velocity relationships for branching actin networks
\cite{Weichsel2010, Schreiber2010}. Likewise, the nature of hysteresis
effects observed in experiment \cite{Parekh2005} remains a subject of
inquiry. Multiple attempts have been made to explain the stall force of
individual actin filaments, but the stall force of a network of cooperating
actin filaments is poorly understood \cite{Mogilner2009a}. Our simulations
yielded a stall force of approximately 2-3 pN/filament (see supplement), which
is in close agreement with the reported value of 1.7$\pm$0.8 pN/filament
\cite{Heinemann2011}. Our reported stall force per filament provides evidence
that actin networks use close to optimal force sharing.

The importance of the number of growing filaments at the boundary determining
how the network responds to load force has previously been suggested for
branching actin networks \cite{Parekh2005, Carlsson2003, Carlsson2001}, and for
bundled filaments such as actin \cite{Tsekouras2011} and microtubules
\cite{Mogilner1999, vanDoorn2000}. In the context of our model, bundled
filaments (actin or microtubule) qualitatively correspond to the case of a very
low branching rate or a high capping rate. According to our calculation,
although the number of contacting filament in this case increases with the load
force, the reinforcement is limited (the lower two curves in Figure
2C). Consequently, the resulting force-velocity curve is always convex,
consistent with the previous findings \cite{Mogilner1999, vanDoorn2000,
  Tsekouras2011}.

Our model therefore suggests that branching event is essential in yielding a
concave force-velocity relationship of actin network growth against load
(Figure 5). It should be noted that a similar conclusion was also reached by
Carlsson's model \cite{Carlsson2003}. Carlsson \cite{Carlsson2003} suggested
that actin networks with autocatalytic branching would continually increase the
number of filaments at the boundary leading to force-independent
velocities. Although branching events are autocatalytic as demonstrated in
experiments \cite{Goley2006a}, Carlsson's model \cite{Carlsson2003} by itself
can only account for the force-insensitive region of force-velocity curves,
i.e., an additional negative feedback would be necessary to limit the increase
in density predicted by the model to produce the inevitably reduction in
velocity at high forces. Moreover, the Carlsson model \cite{Carlsson2003}
predicts a transient hysteresis effect that does not correctly reproduce the
sustained hysteresis observed in experiment \cite{Parekh2005}.

In contrast to Carlsson's model, our model has a built-in negative feedback
mechanism. The capping rate is a first-order reaction of the number of free
filament barbed ends while the branching rate is constant, independent of the
number of filaments. As the number filaments increases with the load force, so
does the capping rate. That effect limits the total number of filaments. As
such, our model explains both the concave and the convex force-velocity curves
without resorting to additional mechanism (Figure 2). Also, our model can
explain the sustained hysteresis in force-velocity relationship
\cite{Parekh2005} (Figure 4(C)).

Our model does not preclude any other negative feedback mechanism limiting the
density of actin filaments. It is likely that increased filament density would
lead to excluded-volume effects at large forces \cite{Schreiber2010}. Any
external negative feedback mechanism would limit the length of the
force-insensitive region of the force-velocity curve. It is important to note
that, due to the exponential term in Eq. \eqref{Brown}, even a relatively small
change in filament density would lead to a large change in velocity. A doubling
of the number of􏰓filaments, $N$ , would lead to an approximately $\displaystyle
\exp\left[\frac{F\delta}{2Nk_BT}\right]$- fold increase in the velocity. A
surprising feature we observed in our simulation was a substantial reduction in
network velocity to extremely small forces ($<$200 pN, see supplemental figure
S3). This reproduces the effect seen in \cite{Parekh2005}, which cannot be
reproduced by \cite{Schreiber2010}. We hypothesize that network velocity has to
be reduced by a sufficient amount before trailing filaments are able to catch
up to the leading edge. When the leading edge is moving close to the growth
rate of individual filaments, trailing filaments are unable to reach the
leading edge. When the leading edge is sufficiently slowed by the
opposing force, trailing filaments are able to reach the leading edge and the
leading edge velocity stabilizes.

It is generally believed that new filament branching in actin networks occurs
in a narrow zone near the membrane \cite{Pollard2000a, Atilgan2005}, although
there is no quantitative measurement on the exact location and the size of such
active Arp2/3 complex-enriched zone. It has been suggested that the restriction
of filament branching at the membrane may have a role in the geometric
organization of the network \cite{Atilgan2005, Schaus2007}. We ran simulations
to test if restricting the area where new filaments could branch would change
the predictions of our simulations. We restricted the branching of new
filaments to a zone of distance L away from, but not at the leading edge, while
keeping all the other model parameters the same as those in Figure 2(B). Here,
the distance L ranges from 2 to 40 actin subunit lengths, corresponding to 5.4
- 108nm. Figure S4 shows that the spatial restriction of branching events (L =
5.4nm - 108nm) does not significantly change the qualitative behavior of
force-velocity relationship as compared to the unrestricted case (Figure
2(B)). If we allowed the branching events strictly at the leading
edge, all the resulting force-velocity curves become convex (Figure S5). As
experiments show that the Arp2/3 complex itself constitutes the first subunit
of the daughter branch \cite{Rouiller2008}, the branching point should be at
least one subunit length away from the leading edge when the first actin
subunit adds to the bound Arp2/3 complex. We therefore deem this case of
branching events strictly at the leading edge may not reflect the
reality. Nonetheless, Figure S4 and S5 suggest that regulation of the active
zone of Arp2/3 complex could modulate the quantitative behavior of the
force-velocity relationship of branching actin network.

The majority of our simulations were performed with constant force. However,
our model is relevant to both constant and non-constant force because it
requires no equilibrium assumptions. Our model showed a relaxation time
($\sim$minutes) before the network reached equilibrium velocity. Changing the
force more slowly than this relaxation time would allow the network to
continuously adapt to the increased forces and show strong hysteresis
effects. Increasing the force substantially faster than the relaxation time
would not allow the network to restructure itself leading to constant force
type results; such phenomena are captured by the convex force-velocity
relationship predicted by our model (see Figure 1(B)). An experimentally
observed value for this relaxation time could be found, and experiments
changing the force pressing against an actin network more slowly than the
observed relaxation time could test our model prediction.

Figure 2(B) should be qualitatively reproducible in experiment and could serve
as an excellent test of several hypotheses for branching actin networks. The
relevant biochemical quantities could be manipulated in the vein of Cameron et
al. \cite{Cameron2004}. Such an experiment could be performed both with
constant force \cite{Marcy2004} and with a constantly increasing force
\cite{Parekh2005}. We further note that the number of active actin nucleation
factor (such as WASP) at the load surface can be negatively regulated by the
number of free barbed ends at the leading edge \cite{Akin2008}. This additional
negative feedback mechanism could limit the effect of increasing the Arp2/3
concentration. Interestingly, capping proteins could increase the growth rate
of branching actin networks by promoting more frequent filament nucleation by
Arp2/3, funneling actin monomers to the uncapped barbed ends of actin
filaments, without affecting the free filament elongation rate
\cite{Akin2008}. These seminal in vitro studies point to a more intertwined
interaction between branching and capping events of actin networks, which will
be the future extension of our current model.

Our model implements a load-sharing mechanism where the contacting filaments
collectively share the load across the leading edge. That is, the addition of a
new actin monomer is only opposed by a fraction of the total load force
pressing on the network. In the context of our model, load-sharing mechanism is
valid as long as Brownian ratchet mechanism holds up. The Brownian ratchet
mechanism assumes that the thermal fluctuations between the filament tip and
the load surface are significantly faster than the addition of new actin
monomers \cite{Mogilner1996, Mogilner2003b}. Fluctuations must be large enough
for a new monomer to fit in the gap between the tip and the load. Smaller and
or slower fluctuations would then reduce the efficiency of the
mechanism. Experiments have indeed demonstrated that reducing thermal
fluctuations by lowering the temperature strongly hinders the efficiency of
filament growth \cite{Shaevitz2007}. Thus, it is the thermal fluctuations that
buffer between the contacting filament tips and the load surface, providing the
flexible interface to accommodate insertions of actin monomers. The separation
of time scales also implies that the load force felt by each contacting
filament is an average over many fluctuations. Consequently, only the partial
load force shared across filaments dictates the network growth rate. In the
future, we intend to study how thermal fluctuations influence the efficiency of
load-sharing mechanism in further detail.

In the model, the actin network was assumed to be a rigid structure so that
does not buckle nor break down. In reality, there will be many cross-linker
proteins that stiffen the actin network. In addition, capping events in our
model limited filaments to an average length of less than 1 $\mu$m,
significantly less than the persistence length of a single actin filament,
$\sim$17 $\mu$m \cite{Ott1993}. Consequently, each individual actin filament
can be viewed as a rigid rod. The actin filaments in our case are highly
branched, which is believed to be much more rigid than its unbranched
counterpart. Although we did not explicitly incorporate these known properties,
our model used their effects and simply assumed that the network was a rigid
structure. The model does not consider the cases where the filaments could
undergo buckling or even breaking down, the topic of which will be investigated
in the future.  

\section*{Conclusion}

The simple physical model shown here gives insight into the behavior of
branching actin network remodeling in the presence of a load. In particular,
the network velocity dependence upon the number of filaments growing against
the leading edge provides a simple mechanical mechanism to explain a number of
experimental effects. The ability of actin networks to remodel is controlled by
the balance between branching and capping rates. This mechanism can account for
both the observed convex and concave force-velocity relationships. Further
investigation into actin network properties, both physical and biochemical,
that determine how many growing filaments a network is able to recruit to the
leading edge will deepen our understanding of actin–based motility.

\section*{Acknowledgements}

We would like to acknowledge Dr. Alex Sodt for his contribution to the code
used for the simulations and Andrea Lively for her helpful comments improving
the manuscript. We also would like to thank Dr. Ed Korn and Dr. John Hammer III
for stimulating discussions and critical suggestions. This work is supported by
the Intramural Research Program of NHLBI at NIH.

\bibliography{actin-references}

\clearpage
\begin{table}
  \caption{Table of model parameters}\label{parameters}
  \begin{tabular}{ c  c  p{0.6\linewidth} }
    \hline
    $\delta$ & 2.7 nm & Length of an actin subunit. \cite{Abraham1999} \\
    $k_BT$ & 4.6 pN$\cdot$nm& Absolute temperature. \\
    $\lambda$ & $200-1,600/s$ & Branching rates.\footnotemark[1] \\
    $\kappa$ & 1-8/s/filament & Capping rates. \cite{Schafer1996} \\
    $v_{free}$ & $100\,\delta$/s & Default velocity. \cite{Abraham1999} \\
    $N_f$ & 200 & Average number of free filament barbed ends.  
    $=\nicefrac{\lambda}{\kappa}$. \\
    $\Theta_{br}$ & $70^\circ$ & Mean branching angle. \cite{Mullins1998} \\
    $\sigma_{br}$ & $5^\circ$ & Branching standard deviation 
    \cite{Mullins1998} \\
    \hline
  \end{tabular}
  \small{$^1$Branching rate was set relative to the capping rate to
    determine the average number of free filaments.}
  \footnotetext[1]{}
\end{table}

\clearpage
\section*{Figure Legends}
\begin{figure}
  \begin{center}
    \includegraphics*[width=3.25in]{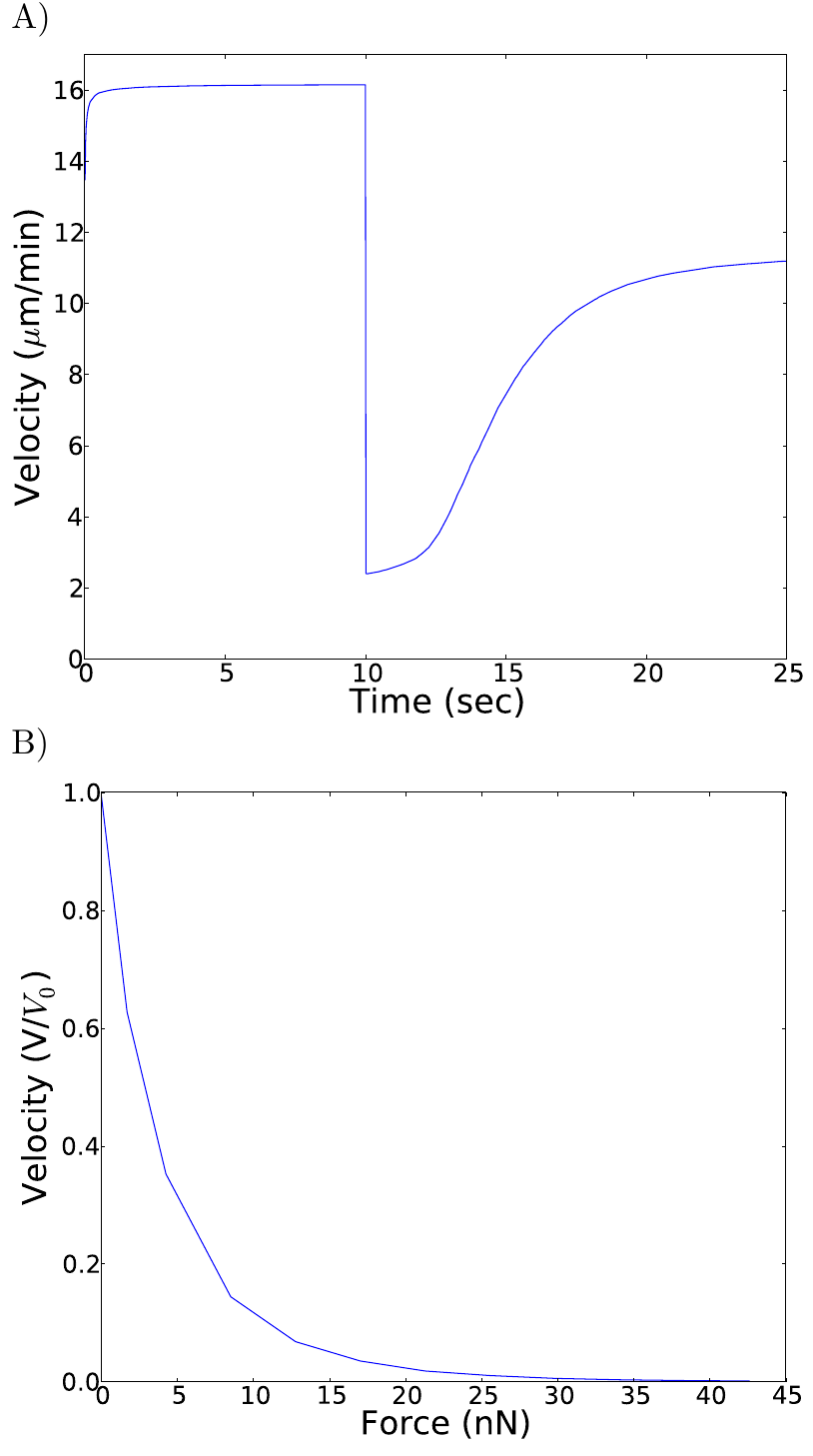}
  \end{center}
  \caption{}\label{short-time}
\end{figure}

\subsection*{Figure~\ref{short-time}}
A) shows a characteristic time trace of the velocity response to force applied
at 100 seconds, and B) shows the convex force-velocity curve generated by the
initial response of the network to force. These simulations were run with 
capping rate $\kappa=1/s/filament$ and branching rate $\lambda=200/s$.

\begin{figure}
  \begin{center}
    \includegraphics*[width=3.25in]{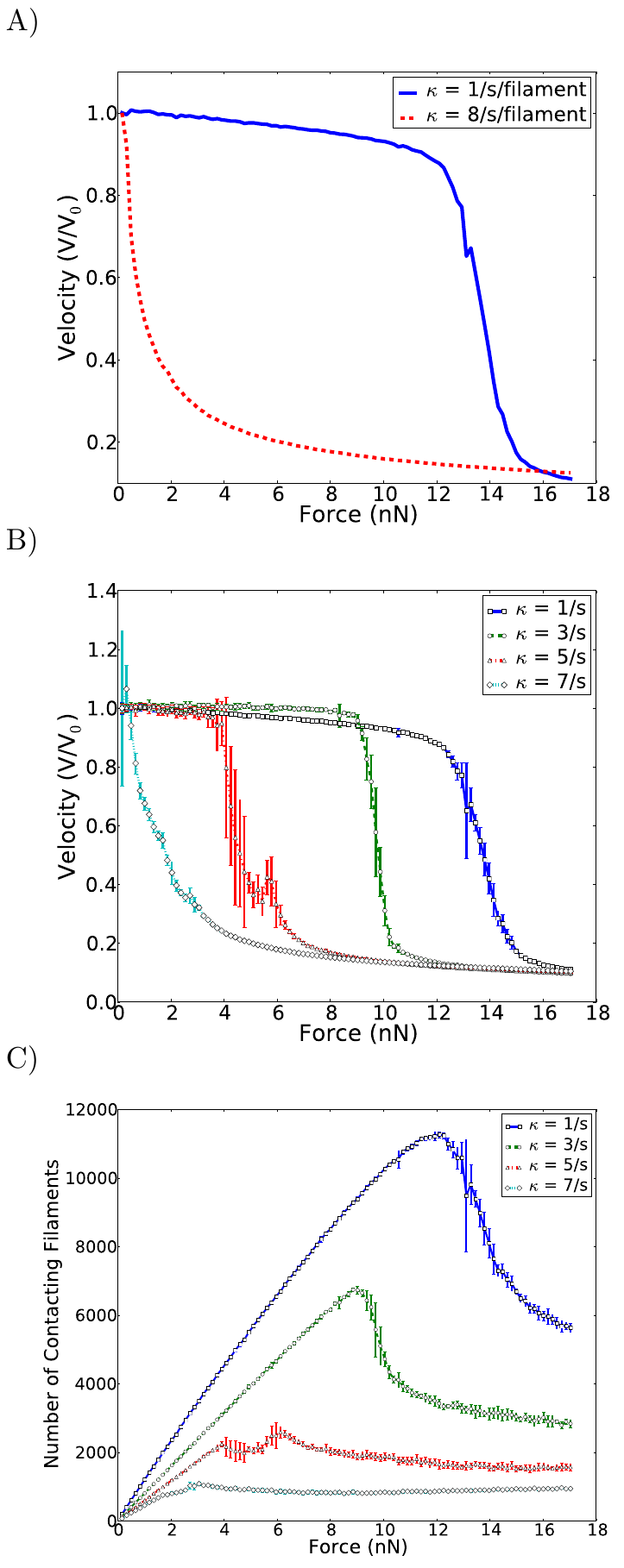}
  \end{center}
  \caption{}\label{continuous}
\end{figure}

\subsection*{Figure~\ref{continuous}}
Varying the capping rate ($\kappa$) over an order of magnitude changes the
shape of the force-velocity curve. A) shows that we are able to qualitatively
reproduce both the convex and the concave force velocity curves. B) shows the
continuous deformation of the force-velocity curve for a few capping rates. C)
shows the relationship between force and the equilibrium number of contacts for
the same capping rates as in B. The error bars in B) and C) represent the
standard deviation estimated from 10 simulations.

\begin{figure}
  \begin{center}
    \includegraphics*[width=3.25in]{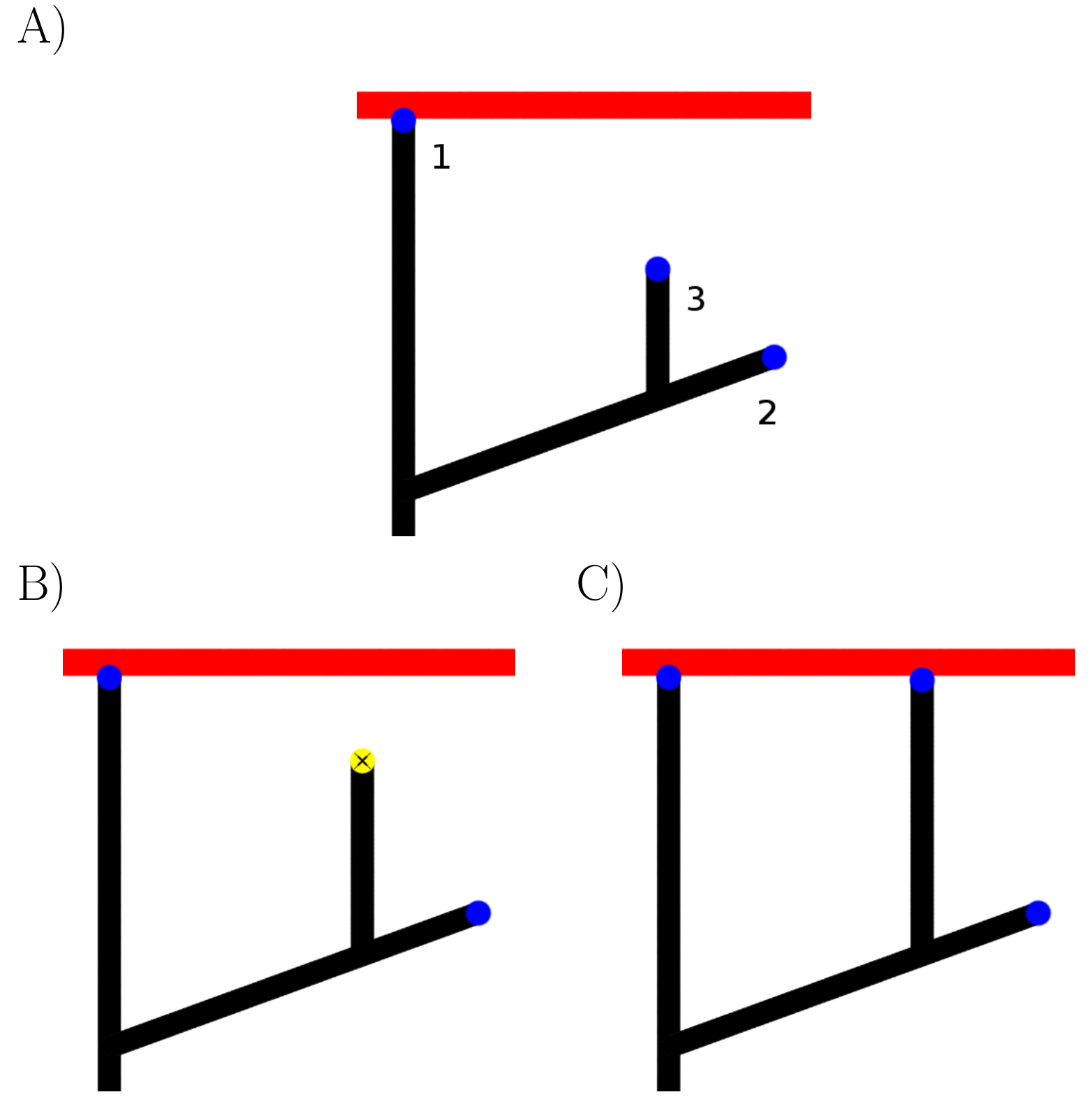}
  \end{center}
  \caption{}\label{diagram}
\end{figure}

\subsection*{Figure~\ref{diagram}}
This diagram shows how the average length of the filaments could influence the
number of filaments in contact with the leading edge. Filaments are represented
by black lines; the red line is the leading edge; the blue circles are actively
growing barbed ends; and the yellow circle represents a barbed end that has
been capped. A) A hypothetical scenario involving one filament barbed
contacting the leading edge with a second filament growing behind the leading
edge. B) When the capping rate is high, the reserve filament is capped (yellow
circle) before it reaches the leading edge. C) When the capping rate is low,
sufficiently long filaments can grow and contact the leading edge, increasing
the leading edge velocity.

\begin{figure}
  \begin{center}
    \includegraphics*[width=3.25in]{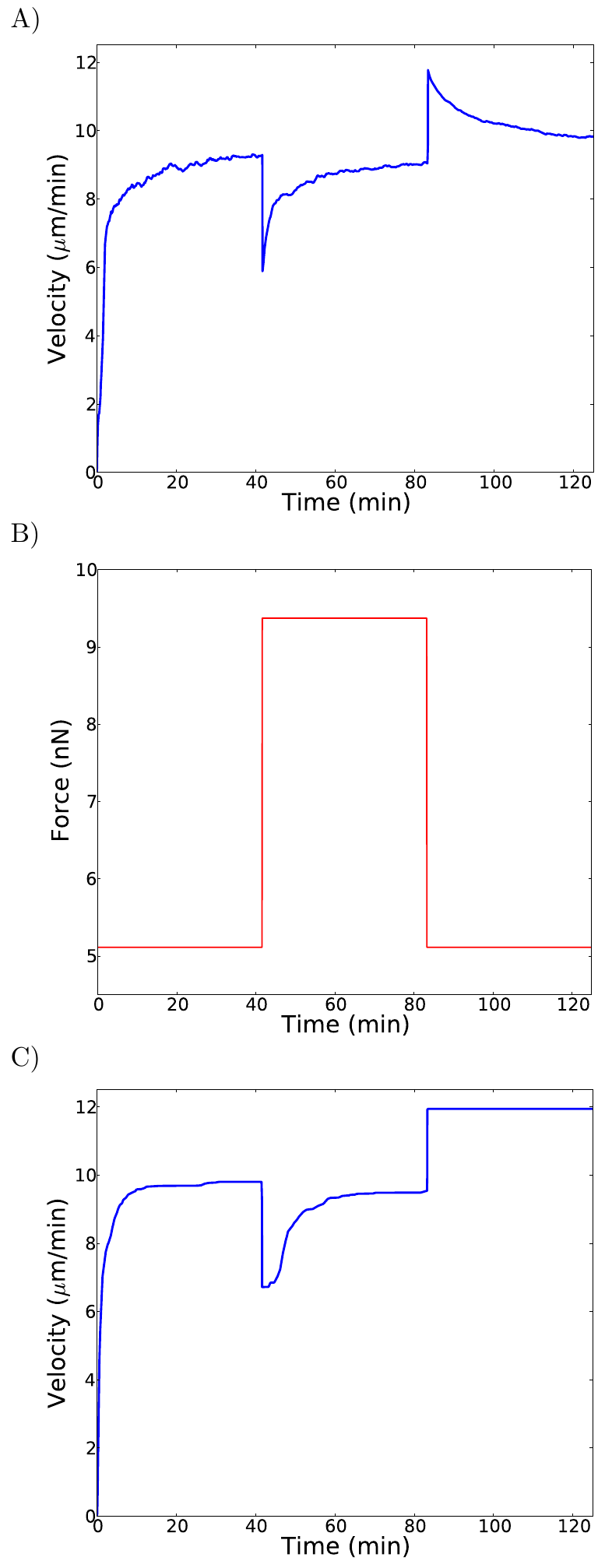}
  \end{center}
  \caption{}\label{reverse}
\end{figure}

\subsection*{Figure~\ref{reverse}}
These are plots of simulations testing for hysteresis in the force-velocity
relationship. A) shows a typical simulation result with a transient hysteresis
effect. B) shows the time dependent force used in both A and C. C) shows a
typical simulation when filaments reaching the leading edge stuck to the
leading edge.

\begin{figure}
  \begin{center}
    \includegraphics*[width=3.25in]{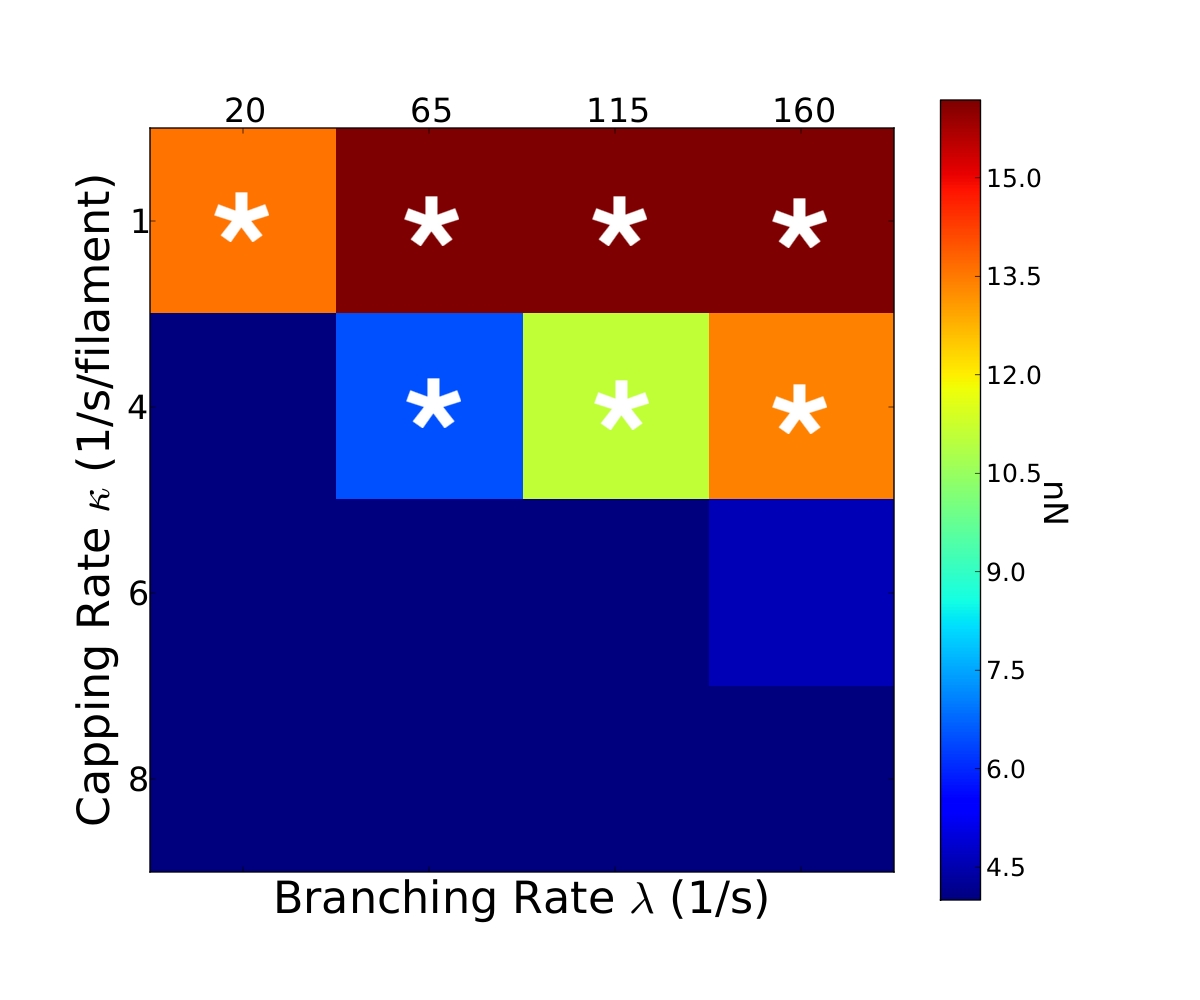}
  \end{center}
  \caption{}\label{phase}
\end{figure}

\subsection*{Figure~\ref{phase}}
The color represents the estimated $f_\frac{1}{2}$ for a variety of branching
and capping rates. The cases with concave force-velocity curves are labeled
with white asterisks. Decreasing the capping rate and increasing the branching
rate serve to generate more filaments which shifts $f_\frac{1}{2}$ rightward,
meaning a more concave curve.

\end{document}


\title{Supplement}
\author{Daniel B. Smith \\ Jian Liu}
\date{}
\maketitle

\section{Equilibration}
The simulations did appear to be sampling from an equilibrium distribution. For
the simulations where the force-velocity relationship was loosely flat, the
velocity converged to an equilibrium quite rapidly, as can be seen in Figure
S1. The curves shown are the average of 10 simulations with the highlighted
region representing the estimated $\pm$ standard deviation.

\begin{figure}[h!]
  \begin{center}
    \begin{minipage}[b]{0.48\linewidth}
      A)

      \begin{center}
        \includegraphics[width=0.9\linewidth]{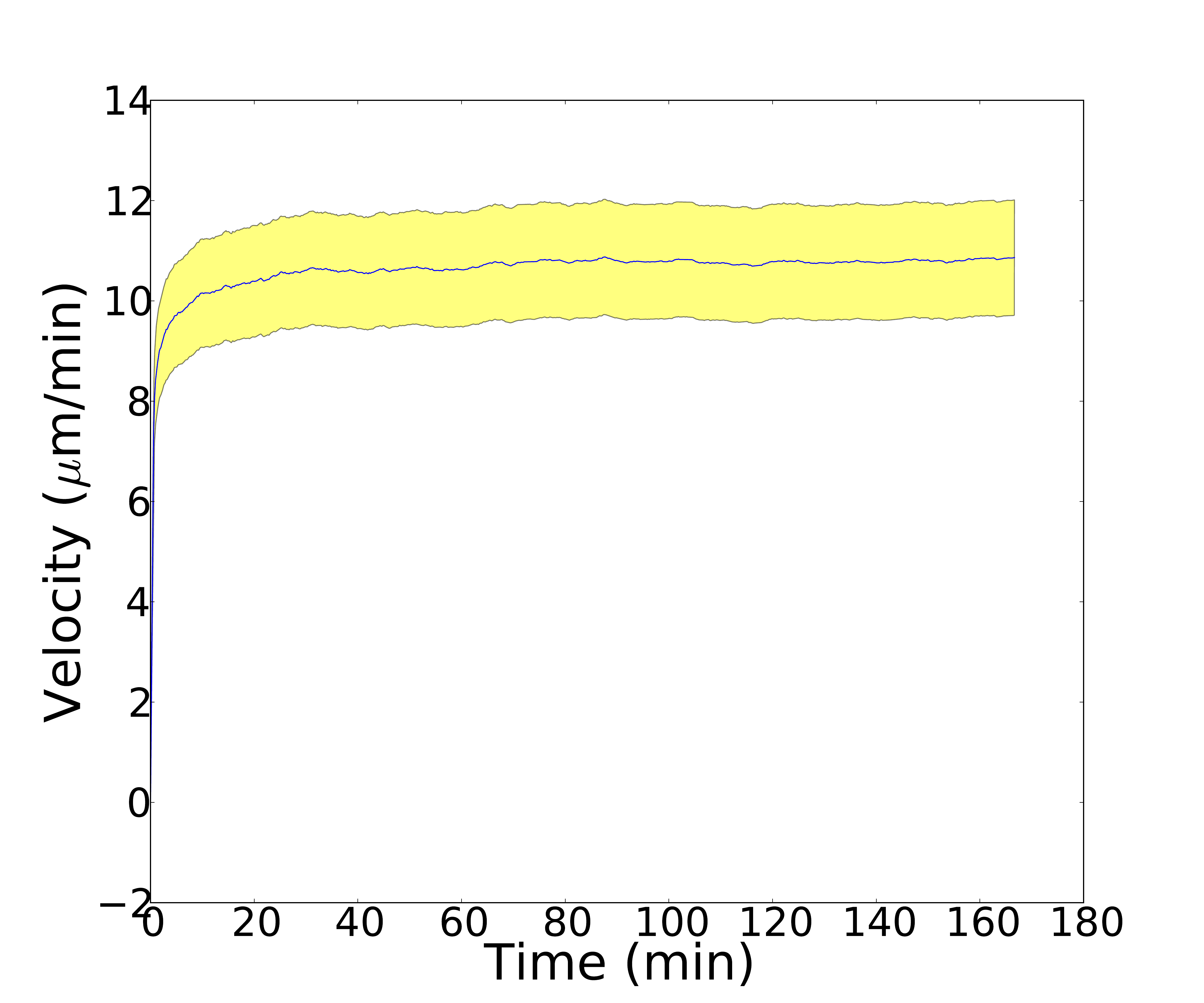}
      \end{center}
    \end{minipage}
    \begin{minipage}[b]{0.48\linewidth}
      B)

      \begin{center}
        \includegraphics[width=0.9\linewidth]{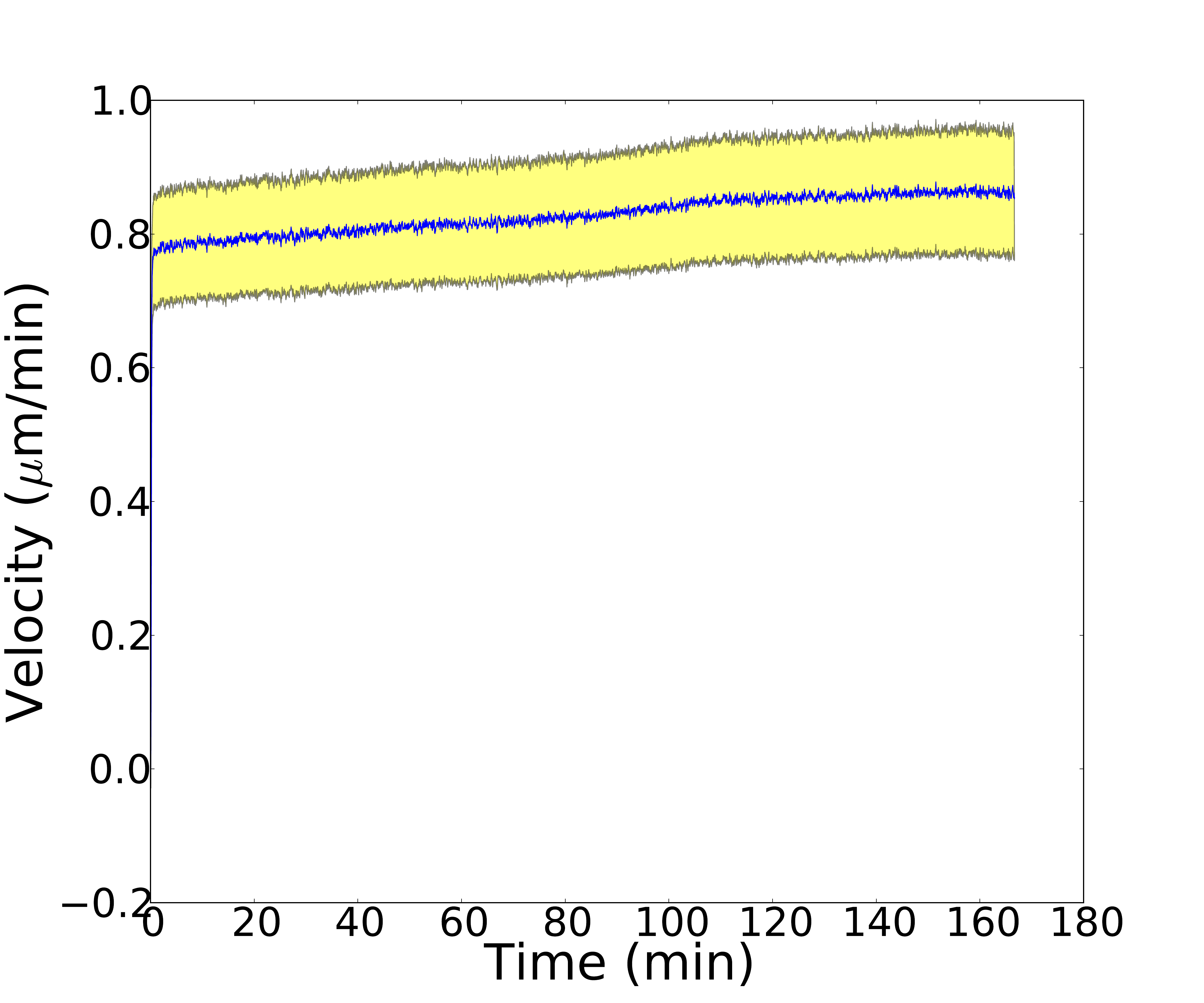}
      \end{center}
    \end{minipage}

    \caption{A) $\kappa$ = 1/s/filament and F = 2.55 nN and B) $\kappa$ =
      3/s/filament and F = 16.2 nN}
  \end{center}
\end{figure}

However, where the force-velocity curve was sharp, the velocity converged only 
at non-physiological time-scales, if at all. This can be seen in Figure S2. 

\begin{figure}[h!]
  \begin{center}
    \begin{minipage}[b]{0.48\linewidth}
      A)

      \begin{center}
        \includegraphics[width=0.9\linewidth]{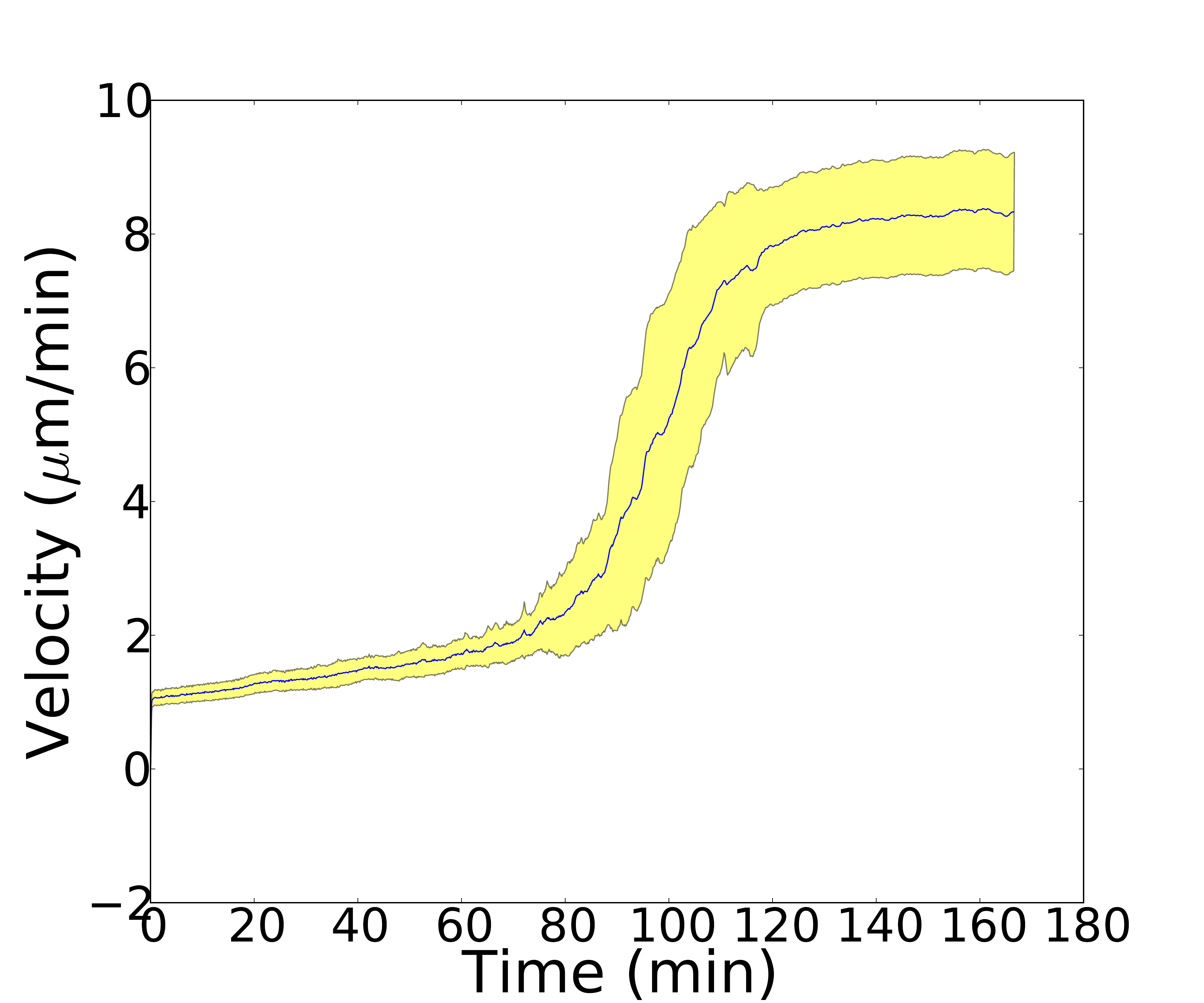}
      \end{center}
    \end{minipage}
    \begin{minipage}[b]{0.48\linewidth}
      B)

      \begin{center}
        \includegraphics[width=0.9\linewidth]{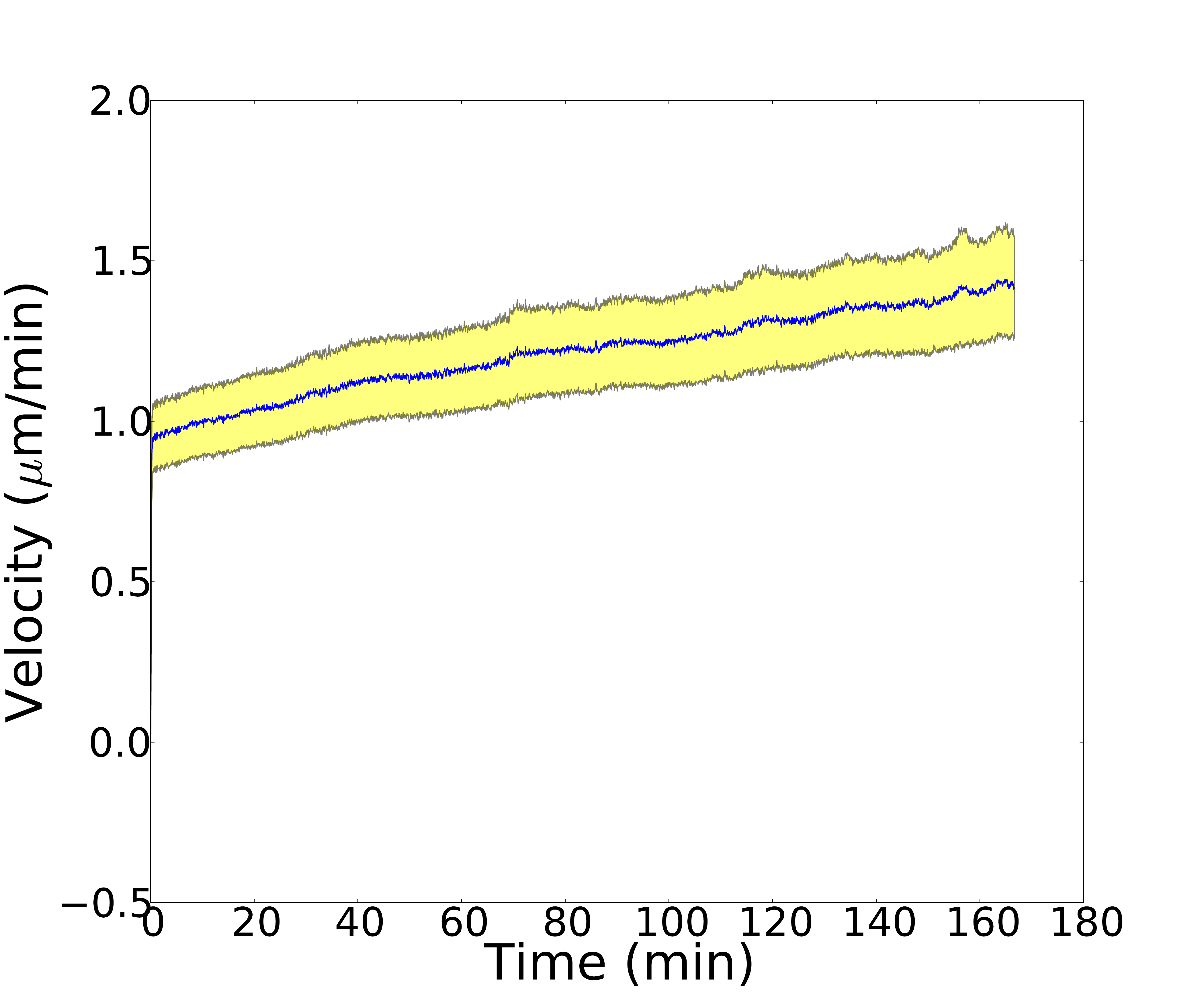}
      \end{center}
    \end{minipage}
    \caption{$\kappa$ = 3/s/filament for both and A) F = 9.4 and B) F = 11.1 nN}
  \end{center}
\end{figure}

\section{Stall Force}
We estimated the stall force per filament by defining the system to be stalled
when an increase in force of 170 pN led to a decrease in velocity of less than
1\%. The stall forces per filament we observed ranged between 0.9 pN and 2.4
pN for $1\leq\kappa\leq5$/s/filament with a mean of 1.3 pN.

\section{Velocity Reduction to Small Forces}
Our model qualitatively reproduces the large velocity reduction in response to
small forces observed in experiment \cite{Prass2006}. The graph below shows the
equilibrium velocity $\displaystyle \nicefrac{v}{v_{free}}$ in response to a
170 pN force. We believe that this is due to the fact that the leading edge
must be sufficiently slowed down for trailing filaments to catch up to the
leading edge. Thus, the force-independent velocities must be significantly
slower than $v_{free}$.

\begin{figure}[h!]
  \begin{center}
    \includegraphics[width=0.6\linewidth]{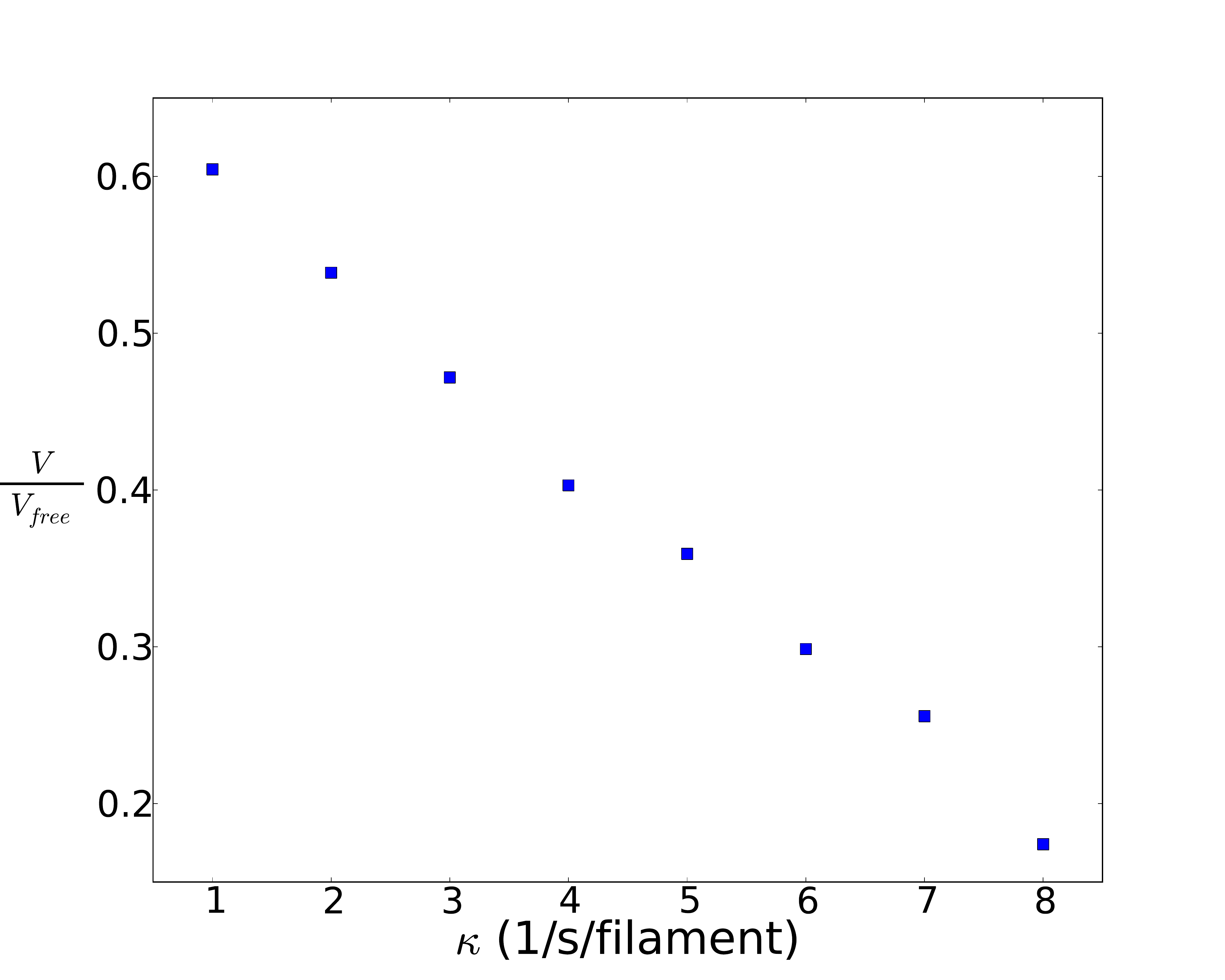}
    \caption{Equilibrium velocity reduction in response to a small ($\sim$170
      pN) force by capping rate.}
  \end{center}
\end{figure}

\section{Spatial Restriction of Branching}

There is experimental evidence that new filaments only branch in a small zone
bordering the membrane. To test how well our results would hold up under that
type of condition, we ran another set of simulations restricting where
filaments could branch. For Figure S4, we kept all of the conditions identical
to simulations in the main text except for that filaments only branched in a
zone of $N\delta$ away from the leading edge.

\begin{figure}[h!]
  \begin{center}
    \includegraphics[width=0.6\linewidth]{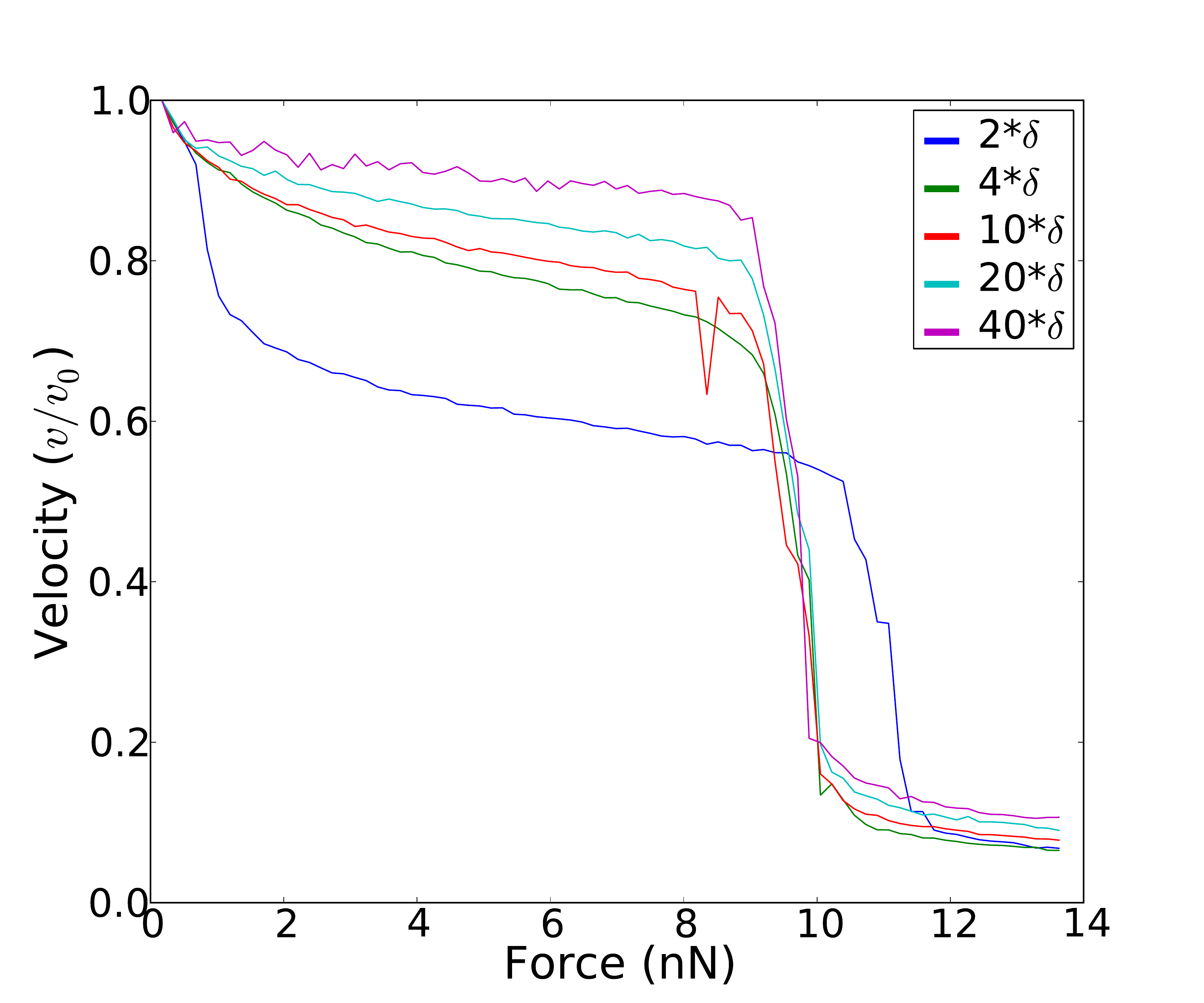}
  \end{center}
  \caption{Force-velocity relationship for $\kappa=3/s/filament$ where
    branching is restricted to $N\delta$.}
\end{figure}

We also tried allowing filaments to branch directly at the leading edge. This
is not necessarily physical as filament tips appear to nucleate new branching
sites \cite{Pantaloni2000}. The results we found are not similar to current 
experiments as can be seen in Figure S5.

\begin{figure}[h!]
  \begin{center}
    \includegraphics[width=0.6\linewidth]{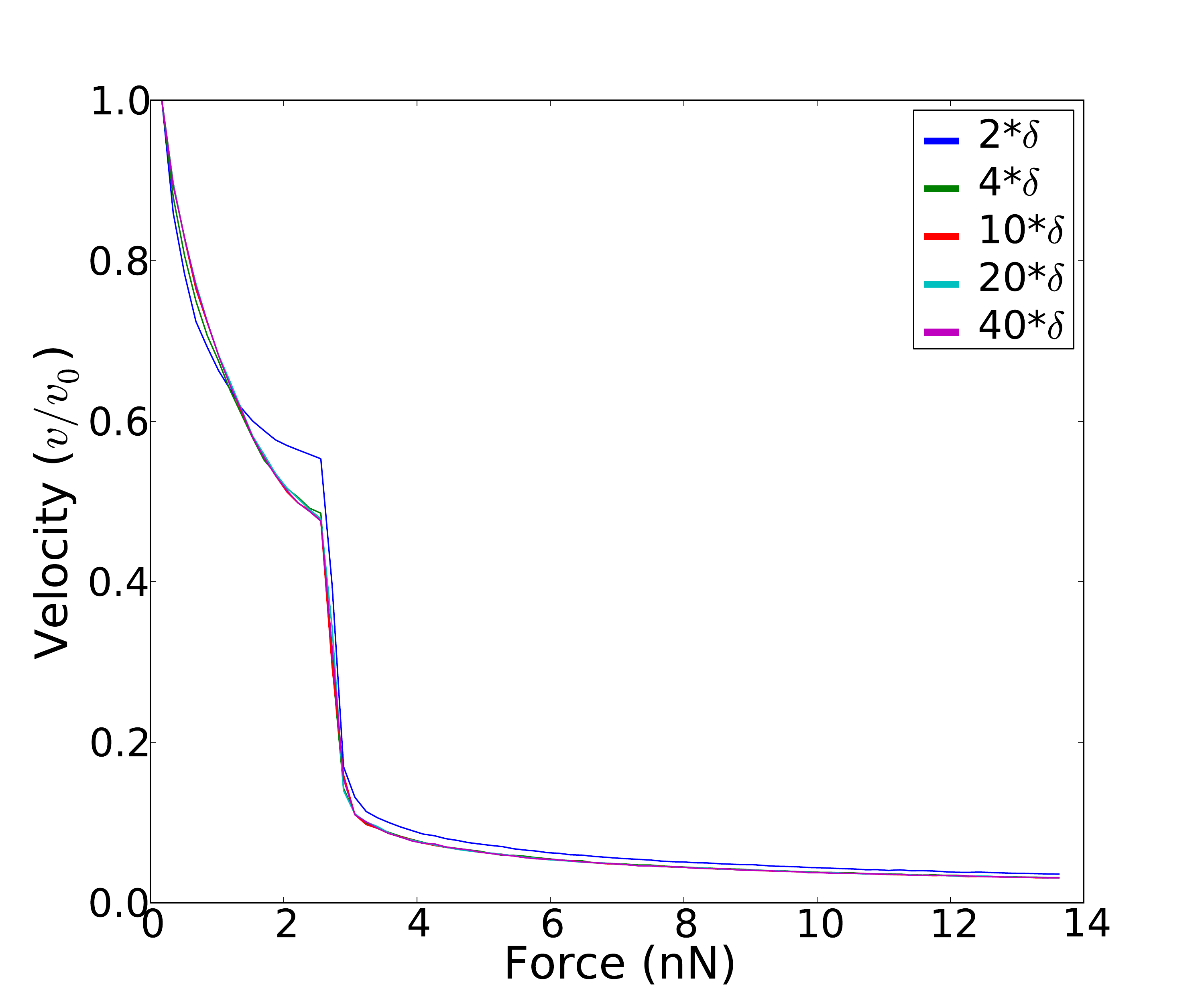}
  \end{center}

  \caption{Force-velocity relationship for $\kappa=1/s/filament$ where branching
    is allowed at the leading edge and restricted to $N\delta$.}
\end{figure}

\clearpage
\bibliographystyle{plain}
\bibliography{actin-references}